\documentclass[
reprint,
superscriptaddress,
amsmath,amssymb,
aps,
pra,
floatfix,
]{revtex4-2}

\usepackage{booktabs}
\usepackage{amsmath,amsthm,amsfonts,amssymb,extarrows}
\usepackage{graphicx}
\usepackage{dcolumn}
\usepackage{bm}

\usepackage{textcomp}
\usepackage{lipsum}
\usepackage[colorlinks=true,citecolor=blue,linkcolor=red,urlcolor=blue]{hyperref}
\usepackage{xcolor}

\begin{document}

\title{Measurement of differential collisional excitation cross sections for the \\K$\alpha$ emission of  He-like oxygen}


\author{Filipe Grilo}\email[]{f.grilo@campus.fct.unl.pt}
\affiliation{Laboratory of Instrumentation, Biomedical Engineering and Radiation Physics (LIBPhys-UNL), Department of Physics, NOVA School of Science and Technology, NOVA University Lisbon, 2829-516 Caparica, Portugal}

\author{Chintan Shah}
\affiliation{NASA/Goddard Space Flight Center, 8800 Greenbelt Rd, Greenbelt, MD 20771, USA}%
\affiliation{Max-Planck-Institut f\"ur Kernphysik, Saupfercheckweg 1, 69117 Heidelberg, Germany}%
\affiliation{Department of Physics and Astronomy, Johns Hopkins University, Baltimore, MD 21218, USA}

\author{Jos\'e Marques}
\affiliation{Laboratório de Instrumentação e Física Experimental de Partículas (LIP), Av. Prof. Gama Pinto 2, 1649-003 Lisboa, Portugal}
\affiliation{Faculdade de Ciências, Universidade de Lisboa, Rua Ernesto de Vasconcelos, Edifício C8, 1749-016 Lisboa, Portugal}

\author{Jos\'e~Paulo~Santos}
\affiliation{Laboratory of Instrumentation, Biomedical Engineering and Radiation Physics (LIBPhys-UNL), Department of Physics, NOVA School of Science and Technology, NOVA University Lisbon, 2829-516 Caparica, Portugal}

\author{Jos\'e R. {Crespo L\'opez-Urrutia}}
\affiliation{Max-Planck-Institut f\"ur Kernphysik, Saupfercheckweg 1, 69117 Heidelberg, Germany}

\author{Pedro Amaro}\email{pdamaro@fct.unl.pt}
\affiliation{Laboratory of Instrumentation, Biomedical Engineering and Radiation Physics (LIBPhys-UNL), Department of Physics, NOVA School of Science and Technology, NOVA University Lisbon, 2829-516 Caparica, Portugal}


\begin{abstract}

We measure the energy-differential cross sections for collisional excitation of the soft X-ray electric-dipole  K$\alpha$ ($x+y+w$) emission from He-like oxygen (O VII), using an electron beam ion trap. Values near their excitation thresholds were extracted from the observed emissivity by rapidly cycling the energy of the exciting electron beam. This allows us to subtract time-dependent contributions of the forbidden $z$-line emission to the multiplet. We develop a time-dependent collisional-radiative model to further demonstrate the method and predict all spectral features. We then compare the extracted $x+y+w$ cross-sections with calculations based on distorted-wave and R-matrix methods from the literature and our own predictions using the Flexible Atomic Code (FAC). All R-matrix results are validated by our measurements of direct and resonant excitation, supporting the use of such state-of-the-art codes for astrophysical and plasma physics diagnostics. 

\end{abstract}

\date{\today} 

\maketitle

\section{Introduction}
\label{intro}
Due to the abundance and stability of He-like ions in a wide range of hot astrophysical plasmas, their bright K$\alpha$ emission lines are essential for the diagnostics of such sources, in particular when high-resolution grating spectrometers onboard  Chandra \cite{Canizares2000,Brinkman2000} and XMM-Newton \cite{Herder2001} are used. 
The simple closed-shell He-like structure favors  $n=2\rightarrow n=1$ transitions, often referred to, following the Gabriel notation \cite{Gabriel1969}, as the resonance line $w$ ($1s 2p~^1\mbox{P}_1- 1s^2~^1\mbox{S}_0$), the intercombination line--composed of two transitions, $x$ ($1s 2p~^3\mbox{P}_2 - 1s^2~^1\mbox{S}_0$) and $y$ ($1s 2p~^3\mbox{P}_1 - 1s^2~^1\mbox{S}_0$)--as well as the forbidden line $z$ ($1s 2s ^3\mbox{S}_1 - 1s^2 ~ ^1\mbox{S}_0$). Their intensity is governed by a combination of atomic processes, such as radiative and dielectronic recombination (RR, DR), collisional ionization (CI), direct excitation (DE), resonant excitation (RE), charge exchange (CX), as well as their subsequent Auger and radiative cascades. Contributions from these processes are differently influenced by macroscopic plasma parameters such as temperature and density, as well as by other conditions, e.~g.,~ being in local thermal equilibrium or a transient state.

Therefore, by studying these lines, astronomers gain insight into the composition and structure of the emitting plasma \cite{Behar2001, Leutenegger2006}. As an example, their intensity ratios are sensitive to the local electron density ($R=z/(x+y)$) and temperature ($G=(x+y+z)/w$) of the plasma (see reviews from Refs. \cite{Kallman2007, Porquet2010, Ezoe2021}), and their Doppler shifts, to plasma velocity \cite{Chung2004, Chen2022}. 
This diagnostic method is applied to a great variety of plasmas: hot stellar coronae \cite{Ness2001, Testa2004, Leutenegger2006, Cohen2022, Gu2022}, accretion disks around pulsars \cite{Schulz2001}; cool active galactic nuclei \cite{Porquet2000,Reynaldi2020}; out-of-equilibrium-plasmas present in the winds of X-ray binaries \cite{Boroson2003, Ignace2019},  supernova remnants \cite{Behar2001, Yamaguchi2009, Uchida2019}, and many extra-solar objects \cite{Liang2006, brickhouse2010, Dubau2002,Orio2020, Henley2005}. 
Ratios of He-like lines are also used for plasma diagnostics in tokamaks for fusion research \cite{Kallne1983, Keenan1989, Bitter2008}.

Before its untimely demise, the Soft X-ray Spectrometer (SXS) microcalorimeter onboard the ill-fated Hitomi satellite acquired high-resolution spectra of the Perseus cluster, allowing to characterize the turbulent motion at its center \cite{Takahashi2016} based on the broadening and Doppler shift of their He-like lines. 
However, differences in theoretical atomic data result in large discrepancies of up to 17\% \cite{Aharonian2018} for the metallicity of Fe derived from the $z/w$ intensity ratio by the spectral codes AtomDB/APEC \cite{Foster2012}, SPEX \cite{Kaastra1996}, and CHIANTI \cite{Young2016}. 
Since this problem arises from differences in the theory sources, experiments are required to assess the quality of the theoretical atomic data. 

Over the last few decades, measurements in electron beam ion traps (EBIT) \cite{Knapp1995,Beiersdorfer1996,Beiersdorfer2002,Shah2021, Nakamura2008,Nakamura2023, Kavanagh2010,Zou2003,Shah2015,Shah2016,Amaro2017, Takashi2017,Tu2017a, Lindroth2020,Orban2024} and ion-storage rings \cite{Linkemann1995a, Hahn1989a, Andersen1989, Orban2008, Hahn2011a, Brandau2012,Bernhardt2014} have been used to benchmark various cross sections of aforementioned line formation mechanisms. In EBIT measurements, the electron-beam energy is swept to extract energy-differential cross sections of various processes, such as DR, RE, and DE. The sweep time is usually a few to tens of milliseconds in order to keep it shorter than ionization and recombination times scales typical for EBITs. This maintains a fairly constant charge-state distribution of the trapped ions during the sweep \cite{Knapp1995,Penetrante1991}. Nonetheless, this method can be affected by the excitation of, and from, long-lived metastable states. In He-like ions, particularly at low-$Z$, the upper state of the line $z$, a magnetic-dipole transition (M1), has a long lifetime $\tau$, in the present case of O VII of $\tau$=($956^{+5}_{-4}$)~$\mu$s \cite{Crespo1998}. If the sweep time is comparable to the lifetime of the transition, retrieving cross sections is challenging since the emissivity due to the excitation of allowed transitions is blended with the radiative decay of the metastable state. This becomes even more difficult when the photon detector resolution is insufficient to resolve the $w$, $x$, $y$, and $z$ lines of K$\alpha$ complex for low-$Z$ ions.

\begin{figure*}
 \centering
\includegraphics[clip=true,width=1.0\textwidth]{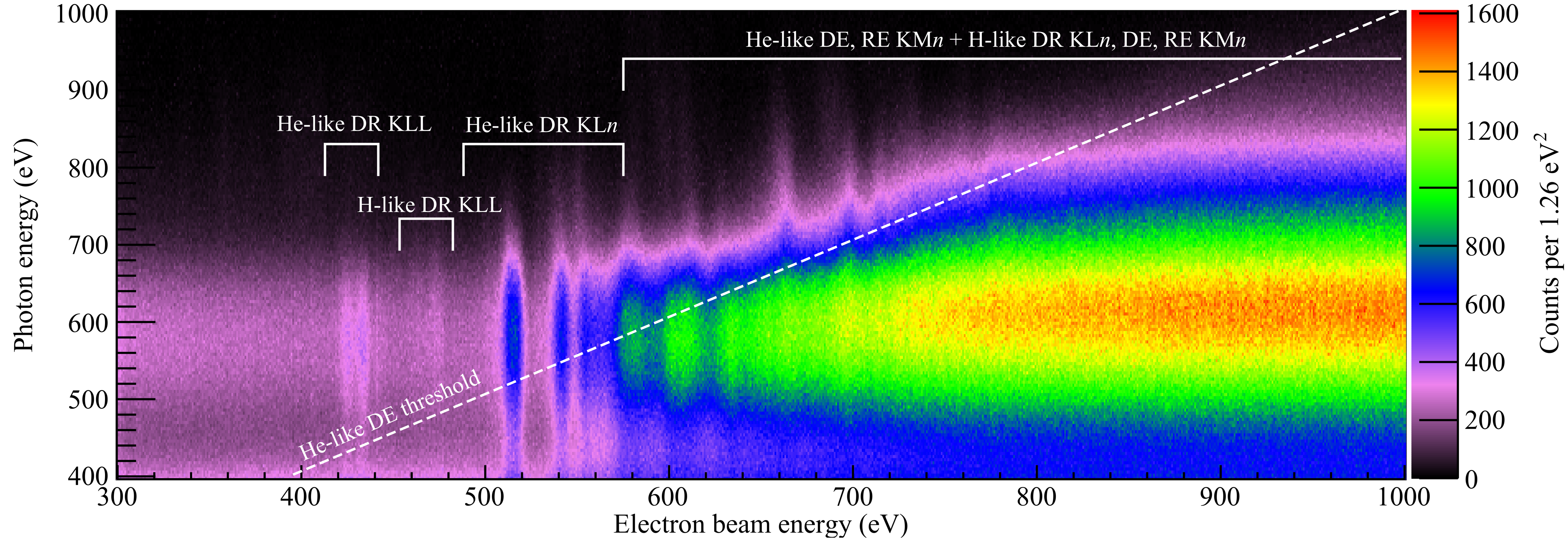}
\caption{Total counts of O VII for both downwards and upwards scans as a function of electron beam and photon energies for the 1 ms scan.}
\label{figs:1}
\end{figure*}

For obtaining the allowed K$\alpha$ cross sections from the overall emissivity, we measured the intensity (counts) of the O VII lines as a function of the electron-beam energy, and varied the sweep rate over the range from 1 ms to 20 ms. This allows us to investigate the time dynamics of the metastable state of the $z$ line, to subtract its contribution from the K$\alpha$ line complex, and to obtain the cross sections for the sum of the $w$, $x$, and $y$ lines. Our predictions using a time-dependent collisional-radiative model agree well with the experimental results. We compare the so-determined collisional excitation cross sections at electron energies near the excitation threshold with our calculations using distorted wave and R-matrix techniques with the Flexible Atomic Code (FAC), as well as with previous calculations and find good agreement between them.

\section{Experiment}
\label{Experi_setup}

\subsection{Setup}

Our measurement was carried out with FLASH-EBIT \cite{Epp2007,Epp2010} at the Max-Planck-Institut f\"ur Kernphysik (MPIK) in Heidelberg, Germany. 
In this device, a mono-energetic electron beam is compressed by a 6 T magnetic field to a diameter of less than 50\,$\mu$m, which dissociates molecular oxygen (O$_2$) injected into the trap region using a tenuous, collimated molecular beam, and sequentially generates by electron impact ionization highly charged ions of this element, with He-like ions being the dominant species. 
The negative space-charge potential caused by the beam strongly confines the ions radially, and a set of biased drift tubes forms an axial, 50 mm long potential well completing an ion trap of cylindrical geometry. 

The soft X-ray emission during the beam energy scans was registered with a silicon-drift detector (SDD) mounted side-on near the trap region, and perpendicular to the electron beam, with resolution around 200\,eV FWHM at 600\,eV, which does not allow us to separate the blend formed by the $z$, $x$, $y$ and $w$ emission lines. However, an excellent electron-beam energy resolution of 5.5\,eV at 520~eV allows us to selectively excite various DR and RE resonances. These often decay by only one transition, which therefore can be effectively separated from others based of selecting a region of interest (ROI) in our data histograms. In such cases, the low-resolution signal from the SDD can yield the transition energy with an accuracy limited by the statistical accuracy of the centroid determination.  
Here we follow a similar experimental scheme based on early studies of DR at the LLNL EBIT \cite{Knapp1989} and later works at MPIK focused on Fe XVII (Ne-like Fe$^{16+}$) \cite{Shah2019, Grilo2021, Grell2024}. 
In all cases, the electron energy scans over a defined region probing the various atomic processes of excitation and recombination leading to photon line emission. Previously, the electron energy duty cycle consisted of a breeding time at a constant energy, followed by a linear ramp-down and a symmetrical ramp-up. The breeding time was necessary for preparing a dominant specific ionic population. Such duty cycle was introduced in Ref.~\cite{Knapp1989, Knapp1993} and used routinely in other measurements of several groups (e.~g., \cite{Yao2010, Xiong2013, Hu2013}).

A periodical triangular duty cycle of range 0.30--1 keV without breeding time was employed to avoid ion losses reducing the count rate that can result from electron beam instablities induced by the fast ramping. Removing the breeding time at high energies for investigating O VII is not a hindrance, since the He-like ionization threshold ($\approx$138~eV) is still below the scanning energies.  
The option of extending the scanning region beyond the H-like ionization threshold of initial $1s^2$ ($\approx$650~eV) was considered for including some amount of O VIII and using previous experimental DR resonances \cite{Kilgus1990} of this ion for the electron-energy calibration. Ionization from the He-like metastable states $1s2s$ is negligible because their populations are lower than $1s^2$ population by a $10^{-5}$ factor (see Sec.~\ref{sec:tdcrm}).  

To probe the time dependence of the emissivity with the forbidden $^3$S$_1$ state (lifetime $\approx0.96~\mu$s), we consider two duty-cycle periods of $20$ ms and $1$ ms. Given its lifetime, $^3$S$_1$ population is exponentially depleted in the $20$ ms scan, while being expected to be almost constant in the $1$ ms one. Due to electronic noise and stability of experimental settings, we did not go below $1$ ms scans. The $z$-line decay is shown in the next Sec.~\ref{sec:metasta} while the overall time-dynamics is investigated in Sec.~\ref{sec:tdcrm}.

\subsection{Metastable state}
\label{sec:metasta}

The measurements resulted in a set of two-dimensional datasets, with counts as a function of electron-beam energy and detected photon energy. For example, the downwards scan for the 1~ms observation is represented in Fig.~\ref{figs:1}. Here, it shows the He-like DR KL$n$ structure converging to the collisional threshold of DE at 560~eV photon energy. The DE continuum is superimposed with resonant structures of RE. DR resonances from H-like O can also be found around 470~eV, and at the beginning of the RE structure of the He-like charge state. The electron-beam energy was calibrated with experimental DR energy centroids from O VIII \cite{Kilgus1990}, while the photon energy was calibrated from previous measurements.

\begin{figure*}
 \centering
\includegraphics[clip=true,width=1.0\linewidth]{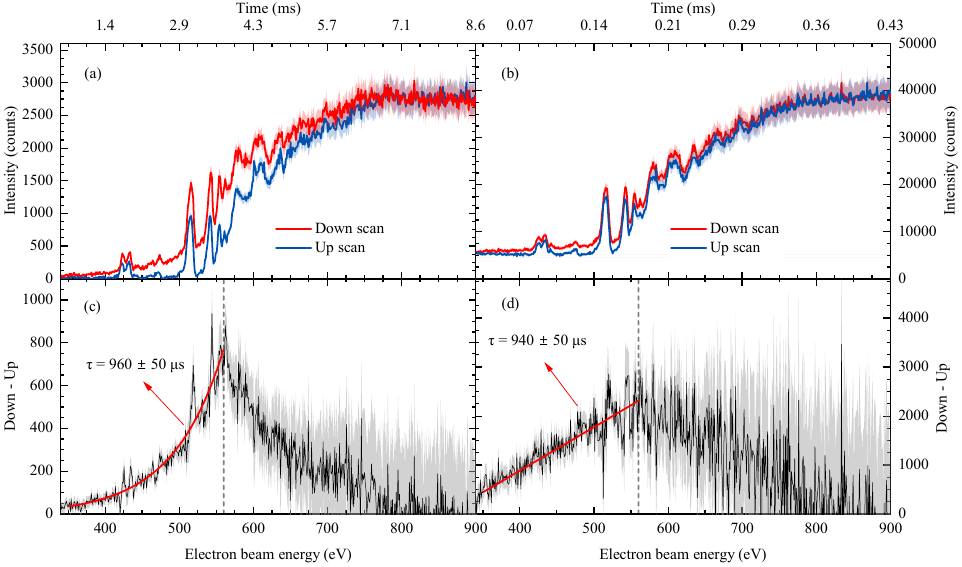}
\caption{Experimental emission from the downwards (red) and upwards (blue) scans as a function of the electron-beam energy for (a) 20 ms and (b) 1 ms cycles. Subplots (c) and (d) contain the subtraction of the emissions (black) from the downward and upward scans, respectively, along with an exponential fit in (c) (red) and, in (d) (red), a fit with Eq.~\eqref{eq:fit}. The dashed vertical lines represent the metastable CE threshold.} 
\label{figs:decay}
\end{figure*}

Top subplots (a) and (b) of Fig.~\ref{figs:decay} show the total photon counts within the region of interest (ROI) compassing 420--720 eV photon energy range, as a function of electron-beam energy. 
Apparent differences arise when comparing the downward scan with the upward scan. The differences between scan directions become more apparent in the 20 ms period (see Fig. \ref{figs:decay} (a)). These differences maybe given by the dynamics of the metastable state $1s2s$ $^{3}S_{1}$, which has an excitation threshold of around 560 eV and a measured lifetime of $\tau$=($956^{+5}_{-4}$)~$\mu$s \cite{Crespo1998}. Given its lifetime, the upward scan of the 20 ms scan is expected to start with virtually no metastable population. As a result, the upwards scan can be used as a baseline for the emissivity without the $z$ line, and the subtraction of the downward scan by the upward scan isolates the emissions of the metastable state in the downwards scan below the threshold energy. Figure \ref{figs:decay} (c) shows the subtraction between scans for the 20~ms case. An exponential fit yielded an apparent ``decay constant" of $0.01494 \pm 0.00006$~$\mbox{eV}^{-1}$, where the uncertainty is given by  fit statistics. Given an electron-beam energy sweep rate of $70\pm4\mbox{~eVs}^{-1}$, we determine a for the metastable state $1s2s\,^3S_1$ a lifetime of $0.96 \pm 0.05$(stats)~ms agreeing with the value reported in Ref.~ \cite{Crespo1998} and calculations therein. 
 Moreover, those differences vanish slowly beyond the excitation threshold, as expected since the metastable population is replenished from there on in both the upward scan and the subsequent ramp down until the electron energy gets again below the threshold. Figure \ref{figs:decay} subplots (b) and (d) show the 1~ms~period scans and their upward-downward differences. Here, we see a similar background at 0.3–0.5~keV in both scans that is not present in the 20~ms scan. For testing if this may be caused by the $z$-line emission from an averaged metastable population fraction, we take the, now much smaller up-down difference and fit  it with the difference of a exponential decay (downward scan) and a time-reversed exponential function (upward scan), given by 
\begin{equation}
D(A,E_i, E_\tau; E)=A \left(e^{(E-E_i)/E_\tau} -  e^{-(E-E_i)/E_\tau}\right),
\label{eq:fit}
\end{equation}
where $A$, $E_i$, and $E_\tau$ are fit parameters related to amplitude, starting energy, and time decay in energy units, respectively. 
The fit yielded a lifetime of $0.94\pm0.05$(stats)~ms, consistent with that of the 20 ms cycle. 
These observations, together with the decay starting at the excitation threshold of that metastable state, prove that it causes the changes between downward and upward scans.
The time evolution of the metastable emissivity is modeled in the next section.

%
\section{Spectral modeling}
\label{sec:tdcrm}
In a previous work \cite{Grilo2021}, we model the emission dynamics of a Fe XVII plasma with a set of population-balance equations for each charge state. Although good results were achieved with this model, it only accounts for the dynamics of the charge-state populations under the assumption that all ions are always excited from the ground state, since the lifetimes of the excited levels are at their longest in tens of microseconds range \cite{Traebert2017,Beiersdorfer2016}. However, for the present O VII measurements, the 1-ms metastable decay delivers a contribution to the emission that needs to be modeled. Furthermore, emissivities obtained from the collisional-radiative model (CRM) package of FAC (FAC-$crm$) \cite{Gu2008} provide steady-state predictions, which have the metastable state depleted below the collisional threshold, i.e., without $z$-line emission. Thus, we implemented a time-dependent CRM to take this effect into account. Following standard balance equations, the state population
$N_{\mathrm{j}}$ can be obtained by numerically solving a set of coupled differential equations given in Ref.~\cite{Penetrante1991},
\begin{equation}
    \begin{aligned}
    \frac{d N_{\mathrm{j}}}{d t}=& \sum_{i}^{\text {states }} n_{e} v_{e}\left[N_{i}\left(\sigma_{i j}^{C E}+\sigma_{i j}^{C I}+\sigma_{i j}^{R R}+\sigma_{i j}^{D C}\right)\right.\\
    &\left.-N_{j}\left(\sigma_{j i}^{C E}+\sigma_{j i}^{C I}+\sigma_{j i}^{R R}+\sigma_{j i}^{D C}\right)\right] \\
    &+N_{i}\left(A_{i j}^{r}+A_{i j}^{a}\right)-N_{j}\left(A_{j i}^{r}+A_{j i}^{a}\right) \\
    & +N_i \sigma_{i j}^{CX} n_0 \Bar{v}_i - N_j \sigma_{j i}^{CX} n_0 \Bar{v}_j, 
    \end{aligned}
    \label{crmEq}
\end{equation}
where $n_e$ and $v_e$ are the electron density and velocity, respectively; $A_{i j}^{x}$ the rates of ($x=r$) radiative and ($x=a$) Auger decays, and $\sigma_{ij}$ the collisional cross section between states $i$ and $j$. DC stands for dielectronic capture (see Sec.~\ref{sec:DW}), and $n_0$ and $\Bar{v}$ the residual gas density and mean ion velocity. Similar studies based in CRM  have been reported elsewhere \cite{Santos2010, Li2015, Kobayashi2015, Ding2020}. All the theoretical energy levels and radiative and nonradiative decay rates were calculated with FAC using the distorted-wave (DW) mode for cross sections following Sec.~\ref{sec:DW}. 
We did not apply R-matrix-based CRM simulations here since it is complex to connect a resonance in the cross section data to an autoionized state with FAC, and thus to address the respective radiative branching ratios of the decay channels. 

\begin{figure*}
 \centering
\includegraphics[clip=true,width=1.0\textwidth]{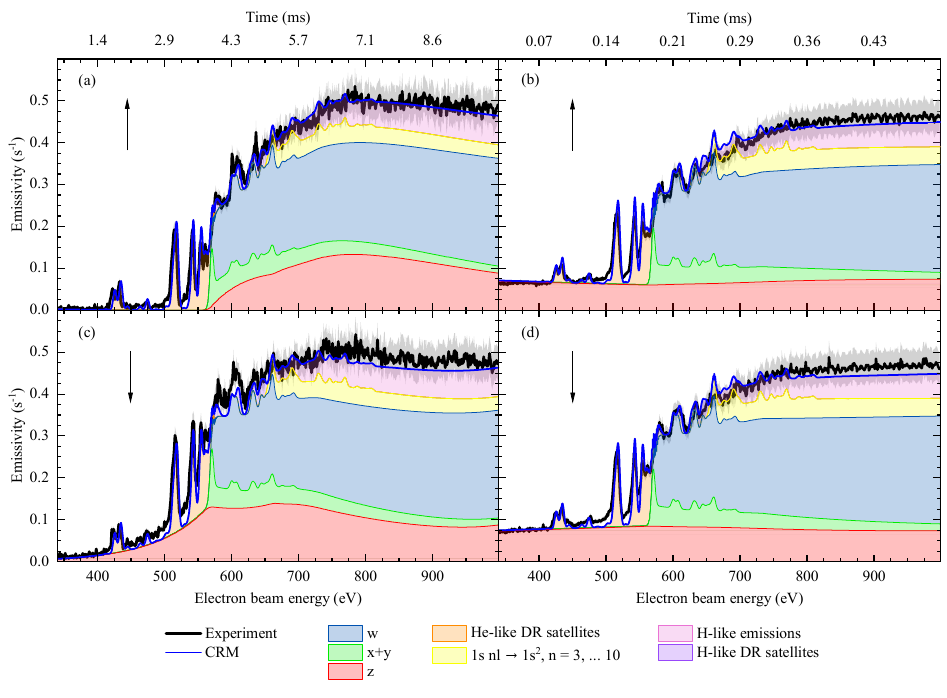}
\caption{Results from the time-dynamical CRM compared to the laboratory measurements under four different experimental conditions: (a) downward scan with a period of 1 ms, (b) downward scan with a period of 20 ms, (c) upward scan with a period of 1 ms, (d) upward scan with a period of 20 ms. Observations are represented in grey with the respective error bars, while the several components of the emission spectrum from CRM are represented in colors. Here we distinguish emissions from the $z$, $x$, $y$, and $w$ lines, $n>2$ excitations from He-like, as well as DR from both H-like and He-like ions.} 
\label{figs:crmResults}
\end{figure*}

The emissivity (defined as photon count rate) of an initial-state population $N_i$ while radiative decaying to a final state $f$ is given by 
\begin{equation}
    I_{if}(E_e) =  N_i A^{r}_{if} , 
    \label{eq:emis}
\end{equation}
with the initial state $i$ being populated by the atomic processes in Eq.~\eqref{crmEq}. Note that if all populations decay by allowed radiative or Auger transition, they quickly reach equilibrium ($\frac{dN_j}{dt}=0$). In such case and not considering CX, the emissivity is approximately given by 
\begin{equation}
    I_{if}(E_e) \approx n_e v_e  N_0 \left(\sigma_{0i} + \sum_k   \sigma_{0k} B_{ki}  \right),
    \label{eq:emis_a}
\end{equation}
where $N_0$ is the ground state population and $B_{ki}$ is the total branching ratio probability of a state $k$ reach the state $i$ after all cascades. The direct link between emissivity and cross sections in Eq.~\eqref{eq:emis_a} has been often assumed in EBIT measurements of cross sections and resonate strengths, e.g., Refs. \cite{Shah2019, Hu2013}. In the present case with a metastable state, extraction of cross sections requires a careful analysis of the emissivity as generally defined in Eq.~\eqref{eq:emis}.   


The simulated emission spectrum is displayed in Fig.~\ref{figs:crmResults} along with the experimental data (top of Fig.~\ref{figs:decay}) adjusted for emissivity. We consider a unitary total population with initial conditions of $N_{1s^2}=0.6$, $N_{1s}=0.3$ and $N_{1s^22s}=0.1$. These populations evolve in the simulation with enough cycles ($>1$~s) of the triangular wave to reach ion equilibrium and a periodical behavior of the $z$-line emission with the cycle. A typical value of electron density of $n_e = 1.0 \times 10^{10} \mbox{cm}^{-3}$ \cite{Grilo2021} was used. For the 1~ms case, final populations in (near) equilibrium are around $N_{1s^2}=0.80$, $N_{1s}=0.18$, $N_{1s^22s}=0.02$ and  $N_{(1s2s) J=1 }= 8\times 10^{-5}$. 
 
The comparison between spectra and predictions highlights that the metastable time dependence explains the differences between upwards and downwards in both scans within the experimental uncertainty. 

In the subplot \ref{figs:crmResults} (a), we can see the 20 ms upwards scan, where there is virtually no \textit{z}-line emission in the lower end energy range of the spectrum after the $^3\mbox{S}_1$ collisional threshold. On the other the subplot \ref{figs:crmResults} (c) shows the 20 ms scan in the downward direction, where an exponential decay of the $^3\mbox{S}_1$ population is evident when the energy crosses below the excitation threshold. After the $^3\mbox{S}_1$ threshold, there are additional decay channels linked to the excitation thresholds of O VII from $1s^2$ to $1s$ $nl$ with $n>2$, where cascades populate the metastable state.

As for the 1~ms scan shown in subplots \ref{figs:crmResults} (b) and (d), these effects are present but more spread over the energy range, leading to the $z$-line forming an almost linear baseline. 
Our CRM was able to reproduce this baseline for both experimental conditions reliably. This gives us confidence that the baseline at the 1~ms emissivity is due to the $z$-line emission and can be safely subtracted allowing direct measurements of the K$\alpha$ ($x+y+w$) cross sections. Nevertheless, some minor discrepancies are observed between the DR resonances (e.g. at 525 eV) that our CRM predictions do not explain. 

This agreement is only achieved with electron densities of $<10^{11} ~\mbox{cm}^{-3}$ and CX Li-like ion rates of $<0.5 ~\mbox{s}^{-1}$ consistent with a Li-like population of $<5$\%. This model prediction is consistent with the absence of Li-like DR resonances in the spectra. This fact might be expected due to the high-compressed electron beam at the 6-T magnetic field and ultrahigh vacuum conditions at the cryogenic FLASH-EBIT.   

 \setlength{\tabcolsep}{6pt}
\renewcommand{\arraystretch}{1.4}
\begin{table}[!t]
\centering
\caption{Error budget of the cross sections at two electron-beam energies (580 eV and 620 eV). The corresponding cross sections and their total uncertainties are also listed.}
\begin{tabular*}{\linewidth}{ @{\extracolsep{\fill}} clcc}
\hline\hline
& Contribution	&	580 eV		&	620 eV		\\	\hline
i & Statistic	and DR fit &	0.6	\%	&	0.6	\%	\\	
ii &ROI	&	3.0	\%	&	2.7	\%	\\	
iii &Detector filter	&	3.0	\%	&	3.0	\%	\\	
iv &H-like DR removal	&	2.4	\%	&	1.0	\%	\\	
v &FAC DR	&	3.9	\%	&	3.9	\%	\\	
&Total	&	6.3	\%	&	5.7	\%	\\	\hline		
&cross section ($\times 10^{-20} \mbox{cm}^{2}$	)&	$5.7 \pm 0.4$	&	$4.8 \pm 0.3$		\\	
\hline\hline
\end{tabular*}\\
\vspace{1ex}
\label{table:2}
\end{table}

\section{Cross section data} 
\label{sec:data_analis}

After the $z$-baselines of the entire spectra were modeled, we focused on the 1~ms scan. This case was selected because of the small baseline variations as a function of the electron-beam energy. The baseline estimated by the CRM was subtracted from the experimental data, resulting in an emission spectrum free of the \textit{z}-line. As expected, both upward and downward spectra coincide after this operation. Hence, both the downward and upward scans were added. Furthermore, the emissivity is proportional to cross sections times electron velocity, following Eq.~\eqref{eq:emis_a}. Therefore, the summed counts were divided by $\sqrt{E_e}$ ($E_e$ is electron energy), resulting in a quantity $I_{\mbox{\small exp}}(E_e)$ to be calibrated to cross sections by a factor.

As in our previous works \cite{Shah2019,Grilo2021}, we normalize the experimental data against the DR-KLL resonant strength of He-like oxygen predicted with FAC, since this channel is the simplest one. For the total uncertainty of the cross sections, we account for various uncertainties as listed in Table \ref{table:2} for two energy cases:  (i) Statistics of each energy bin of the experimental histogram and fit uncertainty of the experimental DR resonances; (ii) choice of the region of interest (ROI) from 420 to 720\,eV photon energy, estimated by comparing a narrow ROI centered on K$\alpha$ versus other ROIs including also K$\beta$ and K$\gamma$ around 665 eV and 695 eV, respectively; (iii) X-ray transmission profile across the ROI of the $1 ~\mu\mbox{m}$ carbon-foil filter blocking UV and optical emissions from the trapped ions in front of the SDD detector, taken as in \cite{Shah2019} from XRCO \footnote{\href{https://henke.lbl.gov/optical_constants/filter2.html}{https://henke.lbl.gov/optical\_constants/filter2.html}}; 
and, (iv) residual DR contributions of H-like O, which are subtracted using data from Ref. \cite{Kilgus1990} with their associated uncertainty. Moreover, (v) we estimate the theoretical normalization uncertainty from the calculations of the He-like DR KLL resonance by increasing their configuration space until approaching numerical convergence and taking the range of calculated DR cross sections between the smallest and largest configuration space as our uncertainty estimate (see Appendix \ref{sec:ap:calc_DR}). 
For our observation at 90\textdegree, we follow Ref.~ \cite{Shah2019} to correct for anisotropic photon angular distribution. A direct measurement of it \cite{Amaro2017,shah2018} showed overall good agreement with FAC predictions. Hence, we also taken here FAC calculations of polarization (FAC-$pol$) for this contribution. Then, the cross section spectra were extracted from the experimental data $I_{\mbox{\small exp}}(E_e)$ by calibrating with 
\begin{equation}
    \sigma_{\mbox{\small exp}}(E_e) =  \frac{S^{\mbox{\small theo}}_{ \mbox{\small KLL}}}{A^{\mbox{\small exp}}_{ \mbox{\small KLL}}}  \frac{ I_{\mbox{\small exp}}(E_e) }{P^{\perp}(E_e)},
\end{equation}
where $P^{\perp}$ is the polarization correction factor for observation at 90\textdegree; $A^{\mbox{\small exp}}_{ \mbox{\small KLL}}$ is the integral of the experimental DR KLL structure, and $S^{\mbox{\small theo}}$ the integrated theoretical cross-sections of the DR KLL. 

As seen in Fig.~\ref{figs:crmResults} (b) or (d), between the DE thresholds of O VII (575~eV) and  O VIII (653~eV), the final spectrum is only composed of the \textit{x}, \textit{y} and \textit{w} lines fed by the resonant structure of RE and continuum DE. Hitherto, predictions for this energy range lacked experimental benchmarks (see next Sec.~\ref{sec:statusquo}).  We review those theoretical works and our FAC calculations for collisional excitation. 

%
\section{Theory}
\label{sec:theory}

\subsection{Theory overview}
\label{sec:statusquo}
First, collisional cross sections were calculated based on the Born approximation \cite{Kim2001} with the Van Regemorter formula \cite{Sampson1992}, giving reasonable results at collisional energies far from the excitation threshold, as reviewd in Ref.~ \cite{Wyngaarden1979}. Distorted-wave (DW) and Coulomb–Born approximations are adequate near the excitation threshold \cite{Elabidi2013, Wu2022, Fritzsche2023}. The resonance-excitation (RE) structure can then be added {\it ad-hoc} in a two-step process \cite{Zhang1987}. In Refs.~\cite{Pradhan1981, Badnell1985}, a DW approach was used for O VII. A Coulomb–Born approximation was applied in Refs.~\cite{Sampson1983, Zhang1987}. General reaction theory of Feshbach \cite{Pindzola1979} and close-coupling approaches, often implemented in R-matrix (RM) formalism \cite{Burke1976, Burke1993}, provide a unified quantum treatment of DE+RE that includes interference between resonance channels and electron continuum. Close-coupling calculations for O VII were introduced in Ref.~\cite{Pindzola1979} with an attached-excited-target approximation, and followed in Ref.~\cite{Kingston1983a, Kingston1983b, Tayal1984} using RM with 6 target $n=3$ autoionizing states. Later, Ref.~\cite{Delahaye2002} included relativistic fine structure with Breit–Pauli RM, radiation damping, and 26 target autoionizing states ($n\le4$). An intermediate-coupling frame transformation (ICFT) implemented in Ref. \cite{Berrington1995, Griffin1998} was also used in Ref.~\cite{Ballance2013, Loch2013} and recently in \cite{Mao2022} to obtain excitation coefficient rates (cross sections averaged over a Maxwellian) of the main lines of O VII. Finally, Ref.~\cite{Aggarwal2008} presents both rate coefficient and cross sections for electron energies at the collisional threshold calculated with the Dirac Atomic R-matrix Code (DARC) and the General Relativistic Atomic Structure Package (GRASP) \cite{Jonsson2007}. Atomic databases that contain coefficient rates from most of these references are XSTAR \cite{Mendoza2021} and the universal atomic DataBase (uaDB) \footnote{\href{https://heasarc.gsfc.nasa.gov/uadb/index.php}{https://heasarc.gsfc.nasa.gov/uadb/index.php}}, which adopt recommended values of Refs. \cite{Kato1989, Zhang1987, Sampson1983}. Others, like AtomDB, CHIANTI, and OPEN-ADAS\footnote{\href{https://open.adas.ac.uk/}{https://open.adas.ac.uk/}} include RM calculations \cite{Tayal1984, Berrington1995, Delahaye2002, Ballance2013}.

\subsection{Present calculations}
\label{sec:calculations}

Theoretical cross sections for DR, DE, and RE were calculated using FAC~\cite{Gu2008}, which provides atomic radial wave functions and the respective eigenvalues from a configuration interaction method with orbitals obtained in a modified electron-electron central potential. This code can calculate cross sections of atomic processes involving a free electron such as CI, DE, and RR both with the DW approximation as well as the R-Matrix (RM) method\cite{Gu2004}.

\subsubsection{Distorted-wave calculations}
\label{sec:DW}

All our FAC-DW calculations included configurations with excited states up to $n=10$. DE for He-like oxygen from the ground and the metastable state to all the 1s $nl$ and 2s $nl$ configurations are included; for the H-like ion we put excitations from the ground state to all the $nl$ configurations, achieving convergence for the energy levels and collisional strengths within less than $0.05\%$ (see Appendix \ref{sec:ap:calc_CE}.)

In the isolated resonance approximation, resonant collisions resulting in DR and RE are approximated by a dielectronic capture (DC), the time-inverse process of Auger, followed by a radiative (DR) or non-radiative decay (RE), but neglecting interference among resonances or with the continuum. Therefore, this two-step process is also known as the isolated resonance approximation. 
Our corresponding DR calculations for the He-like ion include the KL$n$ structure. We proceed analogously for H-like O. As for RE, we focused on the KN$n$ structure.
\subsubsection{R-matrix}
\label{sec:rmat}
The DW approximation is reasonable when only a few resonances are present, but fails when resonances are densely spaced and interfere strongly. In contrast, close-coupling methods take into account quantum interference effects. Thus, for such cases, we use here the R-matrix module of FAC, which is a numerical implementation of the Dirac R-Matrix method, as in the DARC package \cite{Ballance2006}, for the configuration-interaction-based calculations of FAC (see Ref.~\cite{Gu2004} for details). 
We truncate at the $n=6$ level the R-matrix calculations. Their convergence is described in Appendix ~\ref{sec:ap:calc_CE}.
\begin{figure}
 \centering
\includegraphics[clip=true  ,width=1.0\columnwidth]{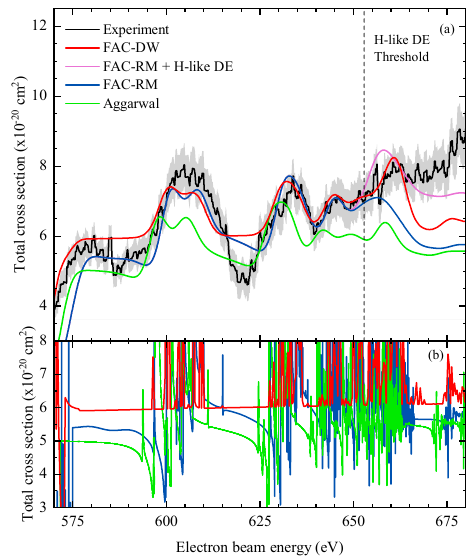}
\caption{Experimental $x+y+w$ cross sections as a function of the electron-beam energy. (a) Experimental results in gray taken from the 1-ms scan, after the \textit{z}line subtraction, are compared with calculations convoluted with a Gaussian according to the experimental resolution. (b) shows unconvoluted theoretical cross sections from Ref. \cite{Aggarwal2008}, FAC DW, and FAC R-matrix, respectively. 
} 
\label{figs:excitation}
\end{figure}
\section{Results and Discussion}
\label{sec:results}
\setlength{\tabcolsep}{6pt}
\renewcommand{\arraystretch}{1.3}
\begin{table*}[!t]
\centering
\caption{Theoretical and experimental cross sections $\sigma$($\times 10^{-20} \mbox{cm}^{2}$) for the collisional excitation from the ground state of the $x$, $y$ and $w$ transitions. 
Collisional energies of 580~eV and 620~eV that are separated from resonances are considered to compare the DE baseline.}
\begin{tabular*}{\textwidth}{ll @{\extracolsep{\fill}} cccccccc}
\hline
\hline

& & \multicolumn{4}{c}{580 eV } &\multicolumn{4}{c}{620 eV } \\
\cline{3-6} \cline{7-10}
&	&	$x$	&	$y$	&	$w$	&	total	theo. & $x$	&	$y$	&	$w$	&	total theo. \\    
\hline
DW & & & & & & & & \\
& Badnell {\it et al} \cite{Badnell1985}	&	&	&	3.886	&		 &		&		&	4.183	&	\\
&FAC-DW	&	1.221	&	0.009	&	4.609	&	5.839$\phantom{00^1}$		&	1.039	&	0.008	&	4.920	&	5.967$\phantom{00^1}$		\\
R-Matrix & & & & & & & & \\
&Tayal {\it et al} \cite{Tayal1984}	&		&		&	3.775	&		&		&		&	4.216	&		\\
&Delahaye {\it et al} \cite{Delahaye2002}	&	1.165	&	0.009	&	3.869	&	5.042$\phantom{00^1}$	&	1.013	&	0,007	&	4.256	&	5.276$\phantom{00^1}$		\\
&Aggarwal {\it et al} \cite{Aggarwal2008}	&	1.131	&	0.008	&	3.818	&	4.958$\phantom{00^1}$	&	0.963	&	0.007	&	4.190	&	5.160$\phantom{00^1}$		\\
&FAC-RM	&	1.211	&	0.009	&	3.969	&	5.189$\phantom{00^1}$	&	1.011	&	0.007	&	4.337	&	5.356$\phantom{00^1}$	\\ \hline
Measurem. & & & &  &  $5.7 \pm 0.4$ & & & & $4.8 \pm 0.3$ \\
\hline
\hline
\end{tabular*}\\
\vspace{1ex}
\label{table:Theo}
\end{table*}

%
\setlength{\tabcolsep}{6pt}
\renewcommand{\arraystretch}{1.4}
\begin{table*}
\centering
\caption{Experimental integrated cross sections ($\times 10^{-20} \mbox{cm}^2$eV) of the lines $x+y+w$ between $575 ~\mbox{eV}$ and $650 ~\mbox{eV}$ compared with calculations made with FAC-DW, FAC R-matrix, and from Ref. \cite{Aggarwal2008}. Values in parentheses are their relative differences to the edge of the experimental uncertainty.}
\begin{tabular*}{\textwidth}{c @{\extracolsep{\fill}} lcccc}
\hline
\hline
	&	FAC-DW		&	FAC R-matrix		&	Aggarwal {\it et al} \cite{Aggarwal2008} 		&	Experiment	\\
\hline
DE	&	441.4		&	403.7	&	375.9	(-1\%)	&	$420 \pm 40$	\\
RE	&	45.5	(-9\%)	&	62.2	(11\%)	&	53.5		&	$53 \pm 3$	\\
Total	&	486.8		&	465.9		&	429.4	(-0.2\%)	&	$470 \pm 40$	\\
\hline
\end{tabular*}\\
\vspace{1ex}
\label{table:int}
\end{table*}

Figure \ref{figs:excitation} (a) shows the excitation cross section near the threshold after the procedure described in Sec.~\ref{sec:data_analis} in comparison with the DARC calculations from Ref.~\cite{Aggarwal2008} and our calculations using FAC, with DW and R-matrix methods following Secs.~\ref{sec:DW} and \ref{sec:rmat}. Below, subplot \ref{figs:excitation} (b) shows the theoretical cross sections not folded with the experimental width. All these calculations reasonably agree with the experiment. 

Concerning the baseline (e.~g.~at 580~eV) often referred to DE, our FAC-DW results are somewhat too high, while previous calculations in Ref.~\cite{Aggarwal2008} slightly underestimate it. Table \ref{table:Theo} compares several calculations of collisional excitation with the measurements at two energies outside RE (resonant) region.  We see a tendency of FAC to yield slightly larger collisional excitation values than the other codes, which might be partially explained by the precision achievable with the configuration-interaction method employed in structure calculations. Moreover, the FAC-DW cross section for the $w$-line is higher than its FAC-RM counterpart, even though it uses similar electronic structure calculations.

We present a quantitative comparison of the RE part for experimental integrated cross sections between 575 and 650 eV in Table \ref{table:int}. The RE and DE parts of the theoretical and experimental values were isolated by assuming that the DE part stays constant over that narrow energy range, as in the DW approach. Although this neglects the interference between resonance and continuum, it yields a resonant contribution that can be compared across all experimental and theoretical sources. 
While the DE values of \cite{Aggarwal2008} seem slightly underestimated, the resonant contributions from that work agree very well with the measurements. As for the total integrated cross sections, all calculations agree with our measurements. 

%
\section{Summary and Conclusions}
\label{sec:sc}

We presented new measurements of the K$\alpha$ emissions of He-like oxygen near the collision excitation threshold, resolving many dielectronic resonances as well as resonant and non-resonant collisional excitation of that transition array. We analyze the time evolution in our experiment in order to account for the sizable contribution of metastables to the emission of the forbidden $z$ transition, allowing us to extract experimental cross sections unaffected by it. Our time-dependent collisional-radiative model reproduces well our experimental conditions. We review available literature on excitation cross-section calculations in comparison with recent DW and RM calculations obtained with FAC. In the $x+y+w$ case, we benchmark recent calculations against our RE and DE experimental data near the excitation threshold. Our results validate those calculations within experimental uncertainties, and support their application to similar species of importance for the diagnostics of hot plasmas in the X-ray region.

%
\vspace{4mm}
\section*{acknowledgments}
Research was funded by the Max Planck Society (MPG), Germany. F.G., P.A. and J. P. M acknowledge support from Funda\c{c}\~{a}o para a Ci\^{e}ncia e Tec\-no\-lo\-gia, Portugal, under grant No .~UID/FIS/04559/2020 (LIBPhys) and contracts UI/BD/151000/2021 and UIDP/50007/2020 (LIP). This research has been carried out under the High-Performance Computing Chair—a R\&D infrastructure (based at the University of Évora; PI: M Avillez), endorsed by Hewlett Packard Enterprise (HPE), and involving a consortium of higher education institutions (University of Algarve, University of Évora, NOVA University of Lisbon, and University of Porto), research centres (CIAC, CIDEHUS, CHRC), enterprises (HPE, ANIET, ASSIMAGRA, Cluster Portugal Mineral Resources, DECSIS, FastCompChem, GeoSense, GEOtek, Health Tech, Starkdata), and public/private organizations (Alentejo Tourism-ERT, KIPT Colab). Financial support was provided by the AHEAD-2020 Project grant agreement 871158 of the European Union’s Horizon 2020 Programme. C.S. acknowledges support from NASA-JHU Cooperative Agreement and MPG. The authors acknowledge the theoretical data provided by K.\,M.\,Aggarwal. 


\bibliographystyle{apsrev4-2}
\bibliography{references_final}

\appendix

\section{Theoretical calculations convergence}
\label{sec:ap:calc}

\subsection{KLL dielectronic recombination }
\label{sec:ap:calc_DR}

\begin{figure*}
 \centering
\includegraphics[clip=true,width=1.0\linewidth]{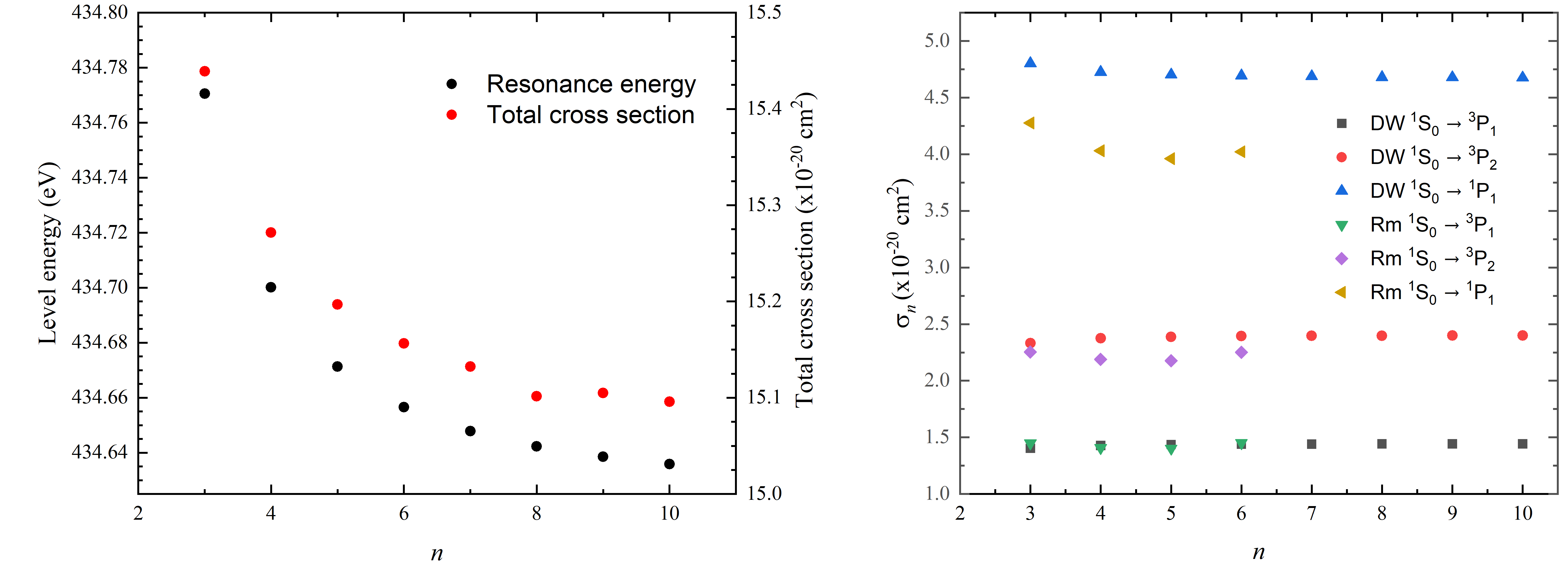}
\caption{(Left) Energy of the most intense resonance of the KLL in function of increments in the maximum principal quantum number $n$. Similar values addressing resonant strength are also represented. (Right) Cross-sections with increments of the maximum principal quantum number $n$ for both DW and  RM methods. } 
\label{figs:convKLL}
\end{figure*}

The FAC DR calculations were made within the isolated resonance approximation. This approximation relies on the values of the resonance energy and the auger and radiative rates from the doubly excited level. The atomic structure for the He-like KL$n$ DR structure was calculated by including the configurations $1sn\ell$ for the He-like ion, $1s2\ell n\ell'$ for the recombined doubly excited state, and $1s^2n\ell$ for the radiative decays. Although the cross-section calibration relied on the KLL structure, with $n=2$, the increase of the principal quantum number enlarges the configuration space included in the configuration interaction calculation, thus improving the accuracy of the energies and decay rates of the KLL structure. Figure \ref{figs:convKLL} (left) shows the convergence of the energy of the most intense resonance of this structure with the maximum allowed principal quantum number $n$ in the structure calculations. A final variation lower than $0.001\%$ was achieved for $n=10$. The convergence of the total cross-section of the KLL structure is also represented, with the few final variations being lower than $0.1 \%$. The variation between the values obtained with a lower and higher number of configurations included in the atomic structure was considered as the uncertainty of the DR structure cross-section.

\subsection{Collisional strength}
\label{sec:ap:calc_CE}

The FAC DW calculations were performed with a He-like atomic structure of the ground and singly excited states with configurations $1$s$^2$ and $1sn\ell$, with $n=2,3,...,10$. 
Collisional excitation calculations made with the DW and RM frameworks sharing the same structure calculations as input were consistent within a few percent. In the present case, that was verified for the excitations from the ground state into the $n=2$ triplet states. The same, however, cannot be said for the $^{1}\mbox{S}_0 \rightarrow  ^{1}\mbox{P}_1$, where a significant difference was found between the two methods, regardless of the configurations included in the structure calculations. Figure \ref{figs:convKLL} (right) shows the conversion between methods for the triplet states excitation, while the singlet state excitation consistently deviates by around $16 \%$ from each other. Comparing the numerical results with values from previous works with different methods (presented in Table \ref{table:Theo}), FAC-DW calculations seem to overestimate the respective collisional strength. Due to the computationally expensive nature of the RM method, it was only possible to perform the calculation with a maximum principal quantum number up to $n=6$.

\end{document}